# Neural encoding of real world face perception

**Authors:** Arish Alreja[1,2,3,4,*], Michael J. Ward[4,5], Lisa S. Parker[6], R. Mark Richardson[5,7,8], Louis-Philippe Morency[9], Taylor J. Abel[2,3,4,10,11], Avniel Singh Ghuman[2,3,4,10]

**Affiliations:**

[1]Machine Learning Department, Carnegie Mellon University, Pittsburgh, USA
[2]Neuroscience Institute, Carnegie Mellon University, Pittsburgh, USA
[3]Center for the Neural Basis of Cognition, Carnegie Mellon University and University of Pittsburgh, Pittsburgh, USA
[4]Department of Neurological Surgery, University of Pittsburgh, Pittsburgh, USA
[5]David Geffen School of Medicine, University of California Los Angeles, Los Angeles, USA
[6]Center for Bioethics & Law, School of Public Health, University of Pittsburgh, Pittsburgh, USA
[7]Department of Neurosurgery, Massachusetts General Hospital, Boston, USA
[8]Harvard Medical School, Boston, USA
[9]Language Technologies Institute, Carnegie Mellon University, Pittsburgh, USA
[10]Brain Institute, University of Pittsburgh, Pittsburgh, USA
[11]Department of Bioengineering, University of Pittsburgh, Pittsburgh, USA

## Abstract

The human brain's ability to comprehend dynamic faces during everyday interactions underpins nuanced social communication in real life, the richness of which cannot be fully captured by controlled laboratory experiments. We captured hours of intracranial brain recordings and wearable eye-tracking from individuals during spontaneous interactions with friends, family, and others. Interpretable machine learning models reconstructed the identity, expressions, and motion of faces participants viewed from brain activity alone. Brain dynamics were predicted solely from viewed faces, highlighting a critical role for the brain's "third visual pathway" for social perception. This pathway encoded facial expressions as deviations from a prototypical neutral expression and exhibited sharper tuning for subtle over intense expressions: a Weber's law for facial expressions, confirmed behaviorally. This work robustly models neural activity during natural social behavior to reveal a non-linear, prototype-centered code particularly finely tuned to critical nuances in facial expressions during rich social perception in real life.

# Main Text

## Introduction

How does your brain encode your daughter's facial expressions and movements while you are playing a board game together (see Movie 1; https://tinyurl.com/4D6F76696531 and video abstract; https://tinyurl.com/6162737472616374 )? This question illustrates a central goal of neuroscience – to understand how the brain processes information during natural behavior in the real world. We study face perception to understand how our brains process the identity, gaze, expressions, and facial movements on people's faces during interactions in real life. Important discoveries, such as the existence of an extended face processing network and aspects of how it codes for faces, have come from laboratory paradigms that monitor brain activity while participants view faces on a screen under experimental constraints[1–9]. However, the fundamental question of how our brains process the expressions and movements of real faces during natural, real world interactions with other people remains open.

Addressing this central neuroscientific goal requires answering two intertwined questions: Can we model the variability of faces during free, natural social interactions in the real world? If so, can we use these models to test hypotheses about neural encoding to understand neural tuning for facial expressions and motion during interactions in the real world? Answering these questions requires a paradigm that not only captures neural data during unconstrained natural behavior but also develops a computational framework to translate the substantial variability of real world stimuli into a structured, analyzable form.

We harnessed simultaneous wearable eye-tracking and intracranial brain recordings from human participants with electrodes implanted in their brains undergoing neural monitoring for seizure localization. This environment afforded a unique opportunity to study high fidelity brain recordings during hours of natural interactions with friends, family, researchers, and others. Participants wore eye-tracking glasses that recorded both their field of view (Fig. 1A) with an outward facing world camera, and where they were looking in the scene with inward facing eye-tracking cameras. Computer vision was used to detect faces in the video from the world camera and combined with eye-tracking to determine when participants looked at faces (Movie 1; https://tinyurl.com/4D6F76696531). The head pose, eye-gaze, shape, texture, expressions, and movement of each face were parameterized using face AI models[10–13]. These models create an interpretable linear space representing faces (Fig. 1C). Brain activity recorded from intracranial electrodes (Fig. 1B) was aligned with eye-movements as fixation-locked brain activity (Fig. 1B) to unravel the neural signatures of the familiar faces of the people participants interacted with.

A defining aspect of the modeling approach is a jointly learned neuro-perceptual space (Fig. 1D) optimized such that each axis of the space corresponds to aspects of brain activity and sets of dynamic facial features that correlate with each other. Moving in this neuro-perceptual space corresponds to both a parametric change in brain activity and complementary parametric changes in the perceptual features that correspond to that brain activity. Neural tuning curves are defined as the relationship between parametric changes in the percept and corresponding parametric changes in brain activity[14–18], and thus, the neuro-perceptual space is a tuning space for facial features. Learning a linear neuro-perceptual space provides straightforward interpretability of the geometry of the data manifold – if the data manifold is linear, there is a simple linear relationship between parametric differences in faces and parametric differences in brain activity. If the data occupy a non-linear manifold with well-defined geometry, this geometry can be examined in the linear space to unravel aspects of faces to which the brain has greater or lesser sensitivity (see supplement for discussion regarding linear versus non-linear approaches for modeling neuro-perceptual relationships).

We first tested the robustness of this approach to model face perception during social interactions by reconstructing faces, including their motion and expressions, from brain activity alone, and then by reconstructing brain activity from faces alone. The jointly learned neuro-perceptual space allowed for bidirectional projections – allowing for both decoding and encoding from a single model. Specifically, to reconstruct a face that a participant looked at, its corresponding fixation locked brain activity was projected into the model's neuro-perceptual space. This neuro-perceptual encoding was then projected out to a face space and the predicted face was visualized (Fig. 1E). This process was then reversed to create a movie of brain activity predicted by the model based on face information alone (Fig. 1E). Subsequently, we examined the tuning geometry to unravel the neural encoding of real world facial expressions and face motion.

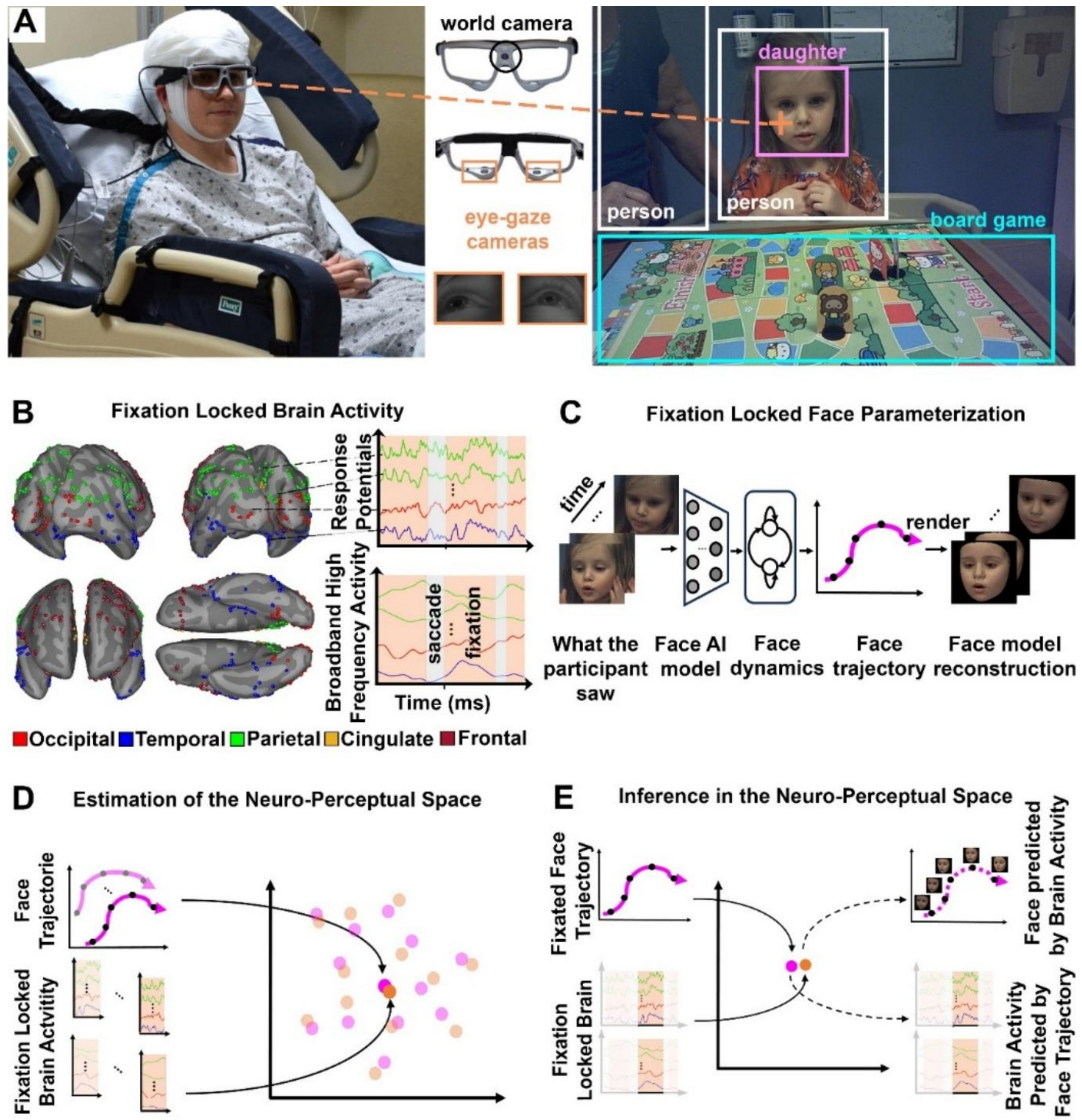

**Fig. 1. (A) Concurrent Brain Activity and Natural Social Interactions:** Participants with intracranial electrodes wore eye-tracking glasses, capturing field of view with a world camera. Inward facing cameras tracked eye-gaze (orange line). Eye-tracking was combined with computer vision annotations of world video, determining when participants viewed faces and who they viewed. **(B) Fixation Locked Brain Activity:** brain activity aligned with eye movements (fixations)was preprocessed to obtain fixation locked potentials and high frequency broadband activity from 687 electrodes distributed across brain areas: temporal (164), parietal (209), occipital (82), cingulate (27), frontal (205). **(C) Face Parameterization:** faces were parameterized to undergo high-fidelity reconstruction using AI models, estimating head pose, eye gaze, identity, expression, and texture in a linear face model and linear dynamical system tracking facial motion. **(D) Learned Neuro-perceptual Space:** a computational model was trained to identify highly correlated brain activity (orange dots) and facial features (magenta dots). **(E) Reciprocal Reconstruction of Faces and Brain activity:** fixation locked activity for held out (test) samples was projected into the neuro-perceptual space (orange dot) and out to the face space, predicting a video of the fixated face. Reversing this process predicted brain activity from a face trajectory (magenta dot) in the neuro-perceptual space.

Recent studies of social perception have led to the hypothesis that there exists a third "middle" visual pathway in addition to the traditional ventral occipito-temporal and dorsal occipito-parietal visual pathways, that is specialized for processing social visual information[9]. This pathway extends superiorly from lateral occipital cortex into the temporal-parietal junction and then anteriorly in lateral temporal cortex down the superior temporal sulcus and superior temporal gyrus. This social vision pathway is particularly sensitive to multidimensional characteristics of faces that are critical to real world face perception during natural interactions, such as face dynamics, expressions, head pose, and eye gaze[19–21], over the unchanging, invariant aspects of faces [9]. However, little is known about how these changeable dimensions of faces, which are particularly relevant to natural social interactions, are coded in the brain. While this study examines the neural basis of real world face perception during natural interactions broadly, we emphasize the social vision pathway due to its theoretical importance to natural social vision and the relative paucity of knowledge regarding its coding principles.

This approach allowed us to reconstruct the faces participants looked at based on brain activity alone and delineate the neurodynamics of real world face perception during natural interactions. It also enabled testing hypotheses about neural encoding and exploration of neural tuning for faces, demonstrating that the social vision pathway encoded dynamic facial expressions as deviations from a prototypical neutral expression. Furthermore, the brain was more sharply tuned for subtle expressions compared to being more coarsely tuned for intense expressions – establishing that Weber's law applies to the neural code and perception of facial expressions.

## Results

### ***Reconstruction of facial features, expressions, and motion from brain activity***

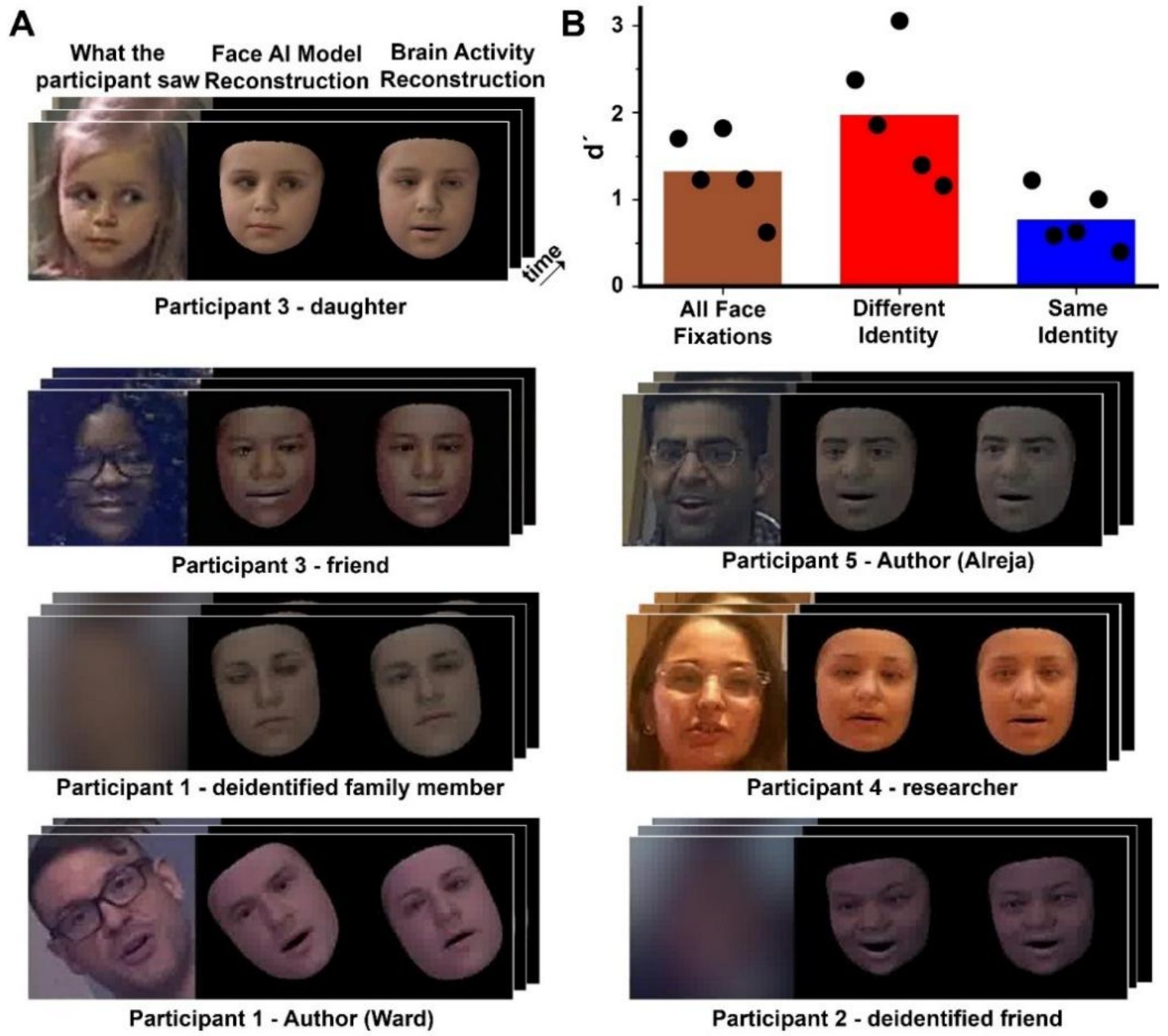

**Fig. 2. (A) Face Reconstruction:** (top left panel) an example face during a fixation (left), its face model representation (middle), and the face reconstructed using brain activity alone (right) for Participant 3's daughter. Faces of multiple individuals were reconstructed for each participant and faces of researchers and clinicians present in recording sessions for different participants could be reconstructed from each participant's brain activity (see Movie 2; https://tinyurl.com/4D6F76696532 for these and additional reconstructions from all participants). The original faces are not shown for individuals who were either unreachable or declined consent to use their faces in print. In those cases, for illustrative purposes in this figure, the face model is rotated by a random matrix that leaves the expressions, motion, and pose unchanged but changes the facial features such that the reconstructed face does not look like the original person. **(B) Top Level Statistics:** In every participant (black dots), d' (an effect size measure on the same standard deviation scale as Cohen's d) was significantly (N=207 to N=2283, $p<0.05$ to $p<0.001$ with permutation tests) above chance (brown; mean d' =1.32), not only between identities (red; mean d' = 1.97), but also between instances of an individual's face (blue; mean d' = 0.77).

Across all participants, accurate movies of the faces that participants looked at could be made using brain activity alone. For example, Fig. 2A and Movie 2; https://tinyurl.com/4D6F76696532 show the face of Participant 3's daughter while playing a board game, as well as additional reconstructed faces from other participants' interactions with friends, family and researchers. To quantify reconstruction quality, we assessed d' by determining if the face reconstruction movie was more like the actual face from that fixation compared to other face fixations. We subdivided these comparisons into comparing all pairs of face fixations, comparing face fixations between different identities, and comparing face fixations restricted to different instances of the same individual's face, i.e., within identity reconstruction of dynamic facial expressions. Significant accuracy (Fig.2B) was observed in each participant for all three comparison types. The qualitative accuracy of individual reconstructions and quantitative statistics demonstrate the robustness of the paradigm, data, and analytical approach for modeling the unconstrained variability of faces in the real world during natural social interactions.

### *Reconstruction of brain activity from dynamic faces*

Fixation locked brain activity could be reconstructed from dynamic facial features alone across several cortical areas (Fig. 3A, Movie 3; https://tinyurl.com/4D6F76696533, Table S1) in the core and extended face processing network[22]. The most robust reconstructions came from electrodes in parietal, temporal and occipital cortex, around the temporal-parietal junction and posterior superior temporal lobe that correspond to the recently proposed third visual stream[9], considered a putative social-vision pathway. These observations were replicated in both hemispheres across different participants (Fig. 3B, Movie 4; https://tinyurl.com/4D6F76696534). The electrodes that had activity for natural facial expressions were significantly correlated to those seen for faces shown on a computer screen in a traditional, "localizer" face paradigm (r=0.21 to 0.34). The correlations imply that the shared variance is statistically significant, though it only accounts for a limited amount of the total variance, highlighting that real world faces during natural social interactions activate a broader network of regions than traditional, controlled face experiments. These reconstructions emphasize activity patterns distributed across the face processing network and highlighted the social vision pathway as critical for face processing in the real world.

### *Neuro-perceptual tuning spaces enable data and hypothesis driven discovery of neural encoding*

The reconstructions of faces from brain activity and vice versa were based on neuro-perceptual spaces learned jointly from brain activity and faces. Neural encoding of dynamic facial expressions during real world interactions can be probed further by examining aspects of the model itself. First, moving along the axes of the neuro-perceptual space enables data-driven discovery of neural tuning. Each axis reveals particular aspects of neural activity and facial features that are related in the model. For example, neuro-perceptual spaces across participants revealed dimensions tuned to particular head poses, identities, expressions, and aspects of facial motion (Fig. 3C, Movie 5; https://tinyurl.com/4D6F76696535). The examples in Movie 5 also show what aspects of brain activity are different for parametric differences in faces. Second, neuro-perceptual spaces can also be used to directly ask questions and test hypotheses about neural tuning for specific aspects of faces, such as identity or expression. Specifically, tuning is revealed by examining how predicted brain activity changes upon moving between points that correspond to different identities or expressions (see supplement for discussion about implications of model generalizability for testing hypotheses about neural encoding). For example, Movie 6; https://tinyurl.com/4D6F76696536 examines how brain activity changes when parametrically shifting between two identities who a participant interacted with in the neuro-perceptual space. Third, the geometry of the data manifold embedded in these tuning spaces can also be used to test hypotheses that reveal neural encoding and tuning for faces, as shown in the next section.

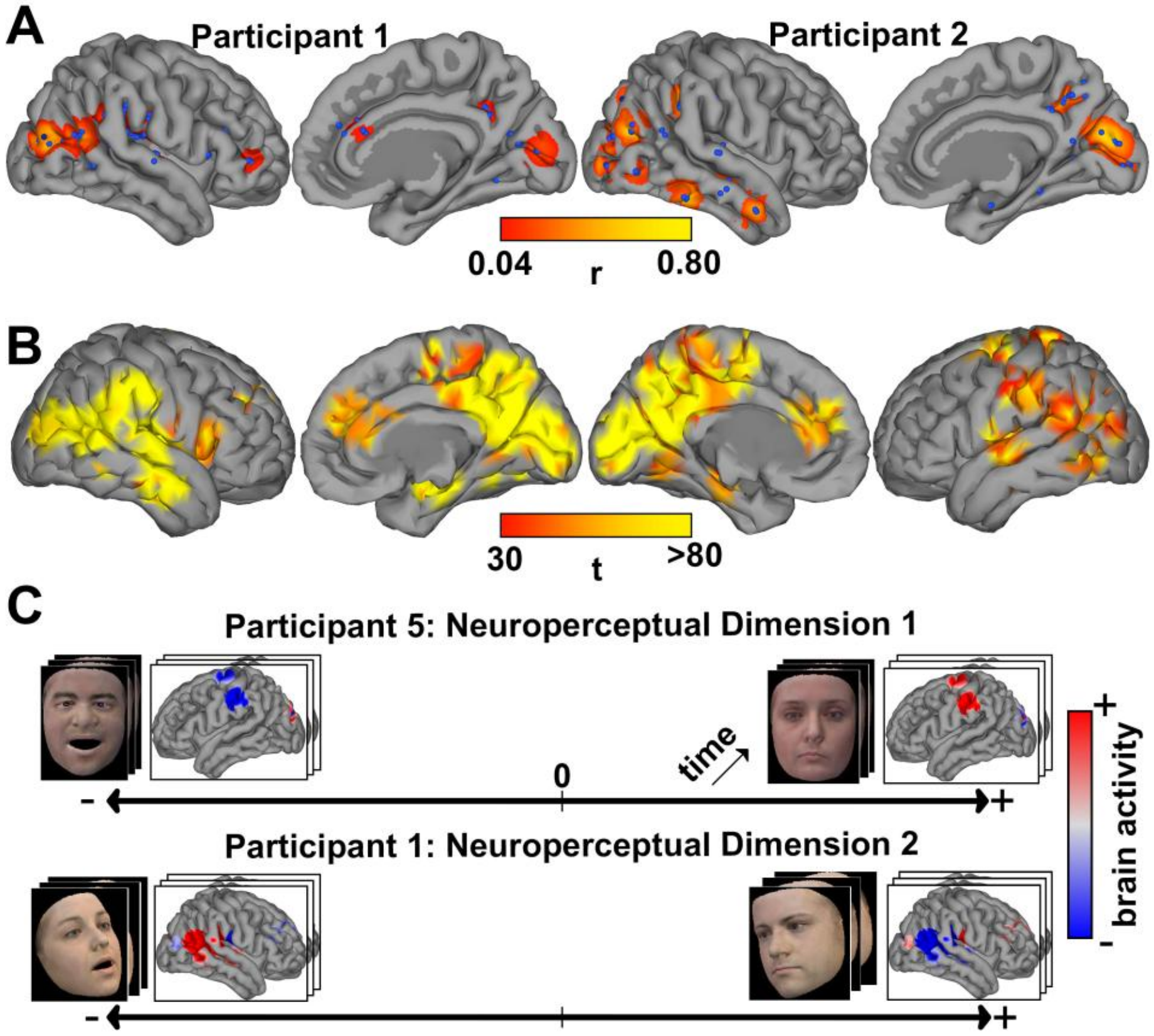

**Fig. 3. (A) Reconstructed Neurodynamics:** Fixation locked brain activity was reconstructed significantly in every participant (N=207 to N= 2283, p<0.01 to p<0.001 across participants; permutation tests) using facial features for electrodes in multiple brain areas (see Table S1 for details), shown here for Participants 1 and 2. Reconstructed neurodynamics are visualized in Movie 3; https://tinyurl.com/4D6F76696533 (including out of sample reconstructions). **(B) Group Analysis of Reconstructed Neurodynamics** Areas of cortex where brain activity was reconstructed significantly at the group level (p < 0.01) based on a mixed effect statistic (also see Movie 4; https://tinyurl.com/4D6F76696534). **(C) Neuro-perceptual Space:** Each axis specifies a mapping between a set of facial features and aspects of brain activity that are significantly correlated with each other. A step along any axis in this space changes the predicted neurodynamics as well as the predicted face, as shown here with examples for Participants 5 (identity) and 1 (sensitive to pose and expression). See Movie 5; https://tinyurl.com/4D6F76696535 for visualization of face and brain dynamics for these and additional examples.

### *Resting faces act as prototypes that anchor cone shaped neural tuning for real world facial expressions*

The geometry of the data manifold in the neuro-perceptual space was examined to investigate how our brains encode variations of someone's face during natural social interactions. Based on theoretical proposals regarding how we code for facial motion[23], we hypothesized that our brains encode facial expressions for a person as deviations from their resting facial appearance and expression, which acts as a prototype for a person's facial expressions. To operationalize this, we first recentered instances of each person who participants saw by estimating the resting face of that person and treating it as the origin of a facial expression space for them. This recentering effectively removes face identity from the model, allowing us to examine how facial expressions are encoded independent of whose face made the expressions. Reconstruction accuracy remained significant even with identity removed (Fig. S1A). In two participants, there were sufficient fixations on multiple faces, and we were able to show that a model trained on one set of people's facial expressions accurately predicted neural responses for other peoples' facial expressions and movements, i.e., cross-identity facial expression reconstruction (Fig. S1B and S1C). In other words, given sufficient data, we could train a model only on person A and person B's facial expressions and movements and accurately predict person C's facial expressions from brain activity (and brain activity from facial expressions), even though the model was not trained with any instances of person C's face. These results validated recentering the face space to remove identity, allowing hypotheses about the coding of face information during real world interactions to be tested.

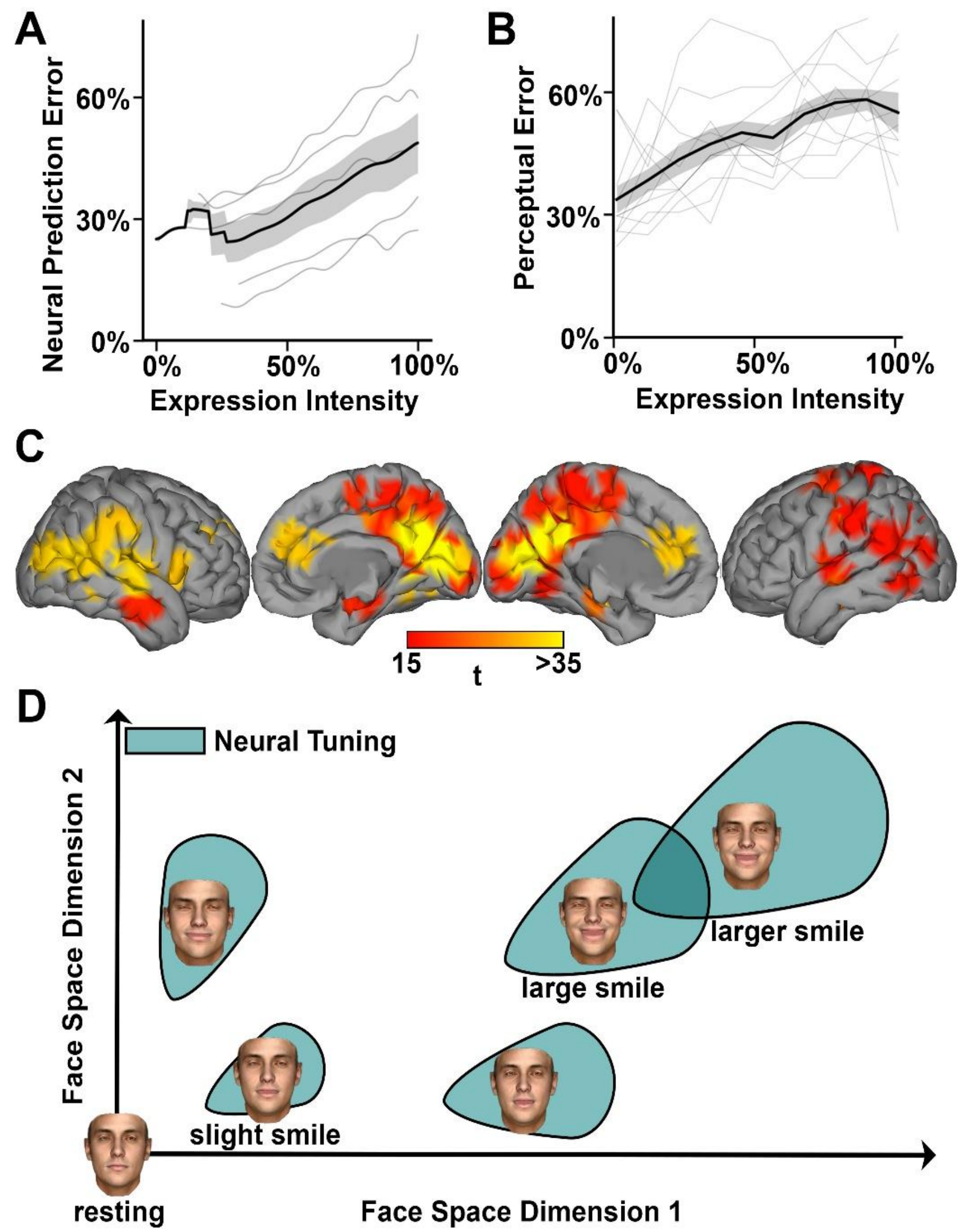

**Fig. 4. An Analog of Weber's Law for Facial Expressions (A)** The (normalized) error of neural prediction increased with the intensity of facial expressions during unscripted natural interactions in the real world (all participants – black, individuals - grey), after corrections to exclude outliers, i.e., regions with too few samples (<15) due to the non-uniform distribution of expression intensity during natural social interactions. **(B)** Behavioral responses in a controlled psychophysical experiment (with uniformly sampled expressions at different intensities) that required participants to discriminate between expressions on faces were consistent with (A), showing decreasing sensitivity for facial expressions as expression intensity increased (all participants – black, individuals - grey). The similarity between the neural and behavioral data supported the interpretation of this finding as a principle of neural tuning, rather than a modeling artifact. **(C) Group Analysis for Weber's Law:** Areas of cortex where neural prediction error was strongly predicted by expression intensity at the group level ($p < 0.01$ assessed by mixed effects statistics (also see Movie 7; https://tinyurl.com/4D6F76696537). **(D) Neural Tuning for Facial Expressions:** The neural and behavioral findings (A, B) suggest cone shaped neural tuning for facial expressions, where neural tuning curves are sharp for subtle expressions, becoming broadly tuned as expression intensity increases i.e., deviations from the resting facial expression become larger (smiles are used as an example for illustrative purposes only). The expression intensity in Fig. 4 is defined in a space of facial features, however the results held in a face defined by visual differences as well (Fig. S2A and S2B). The Weber's law for facial expressions also persisted when facial motion was incorporated (Fig. S2C).

We further hypothesized that the brain's sensitivity to facial expressions would differ based on how far an expression was from a resting expression. Specifically, we hypothesized that our brains should be more sensitive to the difference between a neutral expression and a subtle expression, such as a slight smile, compared to the difference between two more intense expressions that are further away from the resting face, such as a large smile and a slightly larger smile. This hypothesis is based on reasoning that it is critical for us to know what expressions people are making by being sensitive to small differences in expressions around the resting expression, but it is less critical to be able to parse subtle differences in the intensity of a large facial expression. If true, the brain would exhibit a greater error of neural prediction for more intense expressions. Indeed, this is what was found for all participants (Fig. 4A, Table S2). The error in neural prediction corresponded to cortical areas (Fig. 4C and Movie 7; https://tinyurl.com/4D6F76696537) around the temporal-parietal junction and the superior temporal lobe that are part of the recently proposed social vision pathway[9]. Notably, while the results presented in Fig 4A are for static faces derived from the first frame during a fixation, the results remained consistent when face motion over the fixation was incorporated, suggesting that the coding principles are similar for dynamic faces as well (see Fig. S3 and supplemental discussion).

This pattern of sensitivity to expressions as a factor of the distance from a neutral expression is a version of Weber's law[24], instantiated for facial expressions. This law states that "the size of perceptible changes in stimulus intensity is proportional to the intensity of pre-existing stimulus." As an example, we can readily tell the difference between a 2 and 3 pound weight, but find it challenging to tell the difference between a 102 and 103 pound weight. Analogously, the results suggest that our brains can easily differentiate a neutral expression from the Mona Lisa's smile, but find it much harder to tell the difference between a large smile and a slightly larger smile.

The finding of a Weber's law for facial expressions (Fig. 4A) was based on neural responses to real world facial expressions. A critical question is whether the perceptual effects of the observed neural tuning properties are reflected in our behavior. To answer this question, we ran a behavioral psychophysics experiment in a separate group of participants who determined whether expressions on pairs of face images were the same or different. The results of the behavioral experiment demonstrated fewer errors close to the resting expression than further away (Fig. 4B) and confirmed that the neural tuning results reflect our perceptual capabilities. More broadly, we discovered neural tuning properties for facial expressions during uncontrolled real world social interactions and designed a controlled experiment to test hypotheses arising from those observations, highlighting potential synergies and recurrent interactions between real world paradigms and controlled experimentation.

Taken together, these results demonstrate the neural encoding for facial expressions on a person's face during real world interactions is defined by cone shaped tuning (Fig. 4D). The apex of the cone is oriented toward the resting expression. The size of the cone's base, i.e., the width of the underlying tuning curves in the neural population, increases with distance from the resting expression i.e., expression intensity. The increasing sizes of these cones are a geometric interpretation of the neural tuning in the facial expression space, from which a Weber's law for facial expressions emerges (Fig. 4D).

## Discussion

We implemented a framework to understand the neural encoding of faces accounting for the variability of face dynamics and brain activity during real world social interactions between participants and friends, family, researchers, and others. Our approach combined computer vision[10–13] with a dynamical systems model to parameterize facial features, expressions, and motion. A bidirectional model was then used to learn the neuro-perceptual relationships between face parameters and brain activity. This model enabled successful reconstruction of movies of the faces participants looked at based on brain activity alone, and it allowed movies of brain dynamics to be reconstructed based on dynamic facial features alone. The neural reconstructions revealed the importance of activation of regions of the recently proposed third visual stream[9] for social vision in occipital, parietal, and temporal cortex during natural social interactions. The qualitative and quantitative similarity of the reconstructions to the original faces demonstrated the fidelity of this approach. A central feature of the model was the jointly learned neuro-perceptual space, revealing the tuning of neural populations to facial features and enabling testing hypotheses about the neural encoding for dynamic facial expressions. We used this analytical framework to test hypotheses about an anchored encoding scheme where facial expressions on a person's face are centered against their prototypical resting expression. The results supported cone shaped tuning providing an analog of Weber's law for facial expressions, that was subsequently tested and confirmed with a behavioral experiment.

The neural encoding of facial expressions is less understood compared to face identity[2,3]. Real world social interactions, like those recorded here, offer the opportunity to address this gap in knowledge by probing natural variation in facial expressions. Facial expressions, and the motion that underlies these expressions are the changing aspects of faces. The results here supported the idea of an anchored neural representation[23,25], in which dynamic facial expressions are encoded as deviations from the unchanging, invariant prototype of someone's face – their resting facial expression. An important aspect of neural population tuning was a decreasing sensitivity to facial expressions that were farther away from the resting expression – sharper neural tuning for subtle expressions but broader tuning as expressions intensify. In other words, the neural gain is high for subtle expressions and decreases as they become more intense – a Weber's law for facial expressions, which was seen both in the neural and perceptual data (Fig 4A and B). Encoding facial expressions as deviations from a prototype combined with Weber's law provide constraints for computational models of face perception in the brain, and quantifying the magnitude of these effects (Table S2) provides a benchmark for evaluating these models.

Brain activity reconstructed using facial information emphasized the role of the putative social vision pathway[9] within the broader face processing network[22] where brain activity was accurately reconstructed (Fig. 3A, 3B, Movies 3, 4, and Table S1) during natural face perception. This pathway is hypothesized to be important for understanding changing aspects of faces such as facial movements and expressions[23,25], which is supported by accurate reconstructions of face dynamics from neuro-perceptual spaces that emphasized these regions. Cone shaped tuning (Fig. 4D) and a Weber's law (Fig. 4A) for facial expressions provide principles for how the social vision pathway is tuned for faces during real world face processing. These coding principles in the social vision pathway are reflected in how we perceive faces as well (Fig 4B), suggesting that this neural code influences our perceptual behavior.

Reconstructed neurodynamics (Fig. 3A, 3B, Movies 3, 4) and the neural correlates of Weber's law for facial expressions (Fig. 4C and Movie 7) corresponded to a broader network of brain regions than the traditional face processing regions typically considered in laboratory-based face perception experiments. Although brain activity during real world recordings was correlated with controlled experiments, the overlap between them accounted for a limited portion of the overall variation in brain activity, illustrating that faces during real world social interactions activate an extended brain network. Some of the additional brain areas may represent information that was not modeled but was correlated with facial features that were modeled. However, some aspects likely represent a richer neural response associated with processing information about the face of someone in front of you, who you know and are actively interacting with, compared to the type of face processing that occurs when viewing pictures of unknown faces as in traditional laboratory-based experiments[26–30]. Indeed, a common finding from studies in both humans and animals is that a broader set of regions are involved in processing information in natural[31,32] and naturalistic studies[33–40] when compared to similar laboratory experiments. Unraveling the consequences of natural intensity, and the broader activation that accompanies more natural settings, is critical for understanding how the brain processes information during natural behavior.

Although advances in technology enable recording natural behavior at high fidelity, using them to unravel brain-behavior relationships in the real world requires overcoming engineering and analytical challenges that are distinct for each aspect of cognition[37,42–45]. This study illustrates some of the key features for success in using uncontrolled real world recordings to understand neural encoding: tracking appropriate behavioral events for anchoring analysis (fixations in this study), parameterization of stimuli or behavior being related to brain activity (projecting the faces into a parameterized face space), collection of large datasets that transform uncontrolled real world variability from a challenge into an asset (hours of data), and statistical frameworks that robustly reveal the neural underpinnings of perception and behavior (the jointly learned neuro-perceptual space). These themes are also relevant for animal studies that are pushing the boundaries of brain recordings during natural behavior, driven by a rising interest in neuroethology[31,32,39–41,46–48].

Real world neuroscience has the potential to reveal novel insights that can be validated in controlled lab-based experiments, and to interrogate how theory and findings from controlled laboratory based paradigms manifest during natural behavior[41]. However, realizing that potential requires confronting the challenge of modeling uneven and high dimensional variability that arises in truly uncontrolled real-world environments during unscripted natural behavior in a way that enables testing hypotheses to unravel the neural code. This study introduces an interpretable computational framework that models the uncontrolled variability of faces during natural behavior, learning bidirectional neuro-perceptual spaces that enable testing hypotheses. The results reveal a picture of neural encoding and tuning for faces that features a non-linear, prototype anchored encoding rule and cone shaped tuning that establishes a Weber's law for dynamic facial expressions during natural face perception in unscripted, real world social interactions. Taken together, these findings demonstrate how the promise of real world neuroscience can be realized by tackling distinct modeling and inferential challenges to test hypotheses, probe neural tuning properties of the brain in natural settings and make new discoveries in the most intrinsically ecologically valid setting, real life.

# Materials & Methods

## Participants

A total of five participants (three male, two female; per clinical records) underwent surgical placement of intracranial EEG (iEEG) depth electrodes as standard of care for seizure zone localization. The ages of the participants ranged from 16 to 64 years old (mean = 35.4 years, SD = 16.3 years). No seizure events were observed during experimental sessions.

## Informed Consent

All participants provided written informed consent in accordance with the University of Pittsburgh Institutional Review Board. The informed consent protocols were developed in consultation with a bioethicist (Dr. Lisa Parker) and approved by the Institutional Review Board of the University of Pittsburgh. Audio and video of personal interactions were recorded during experimental sessions. Our protocol incorporated several measures to ensure privacy considerations and concerns could be addressed based on the preferences of individual participants. First, the timing of recording sessions was chosen based on clinical considerations and participant preference to ensure that they were comfortable with recording of their interactions with the visitors present (and/or expected to be present). Second, all visitors present in the room were notified about the nature of the experiment at the beginning of each recording session and given the opportunity to avoid participation. Third, a notification was posted at the entrance of the participant's hospital room informing any entrants that an experiment was being conducted in which they might be recorded so that they could choose whether to be present or not. Finally, at the end of each experimental recording, participants were polled to confirm their consent with the recording being used for research purposes and offered the option to have specific portions (e.g., a personal conversation) or the entire recording deleted if they wished. Thus, explicit "ongoing consent" was acquired at the beginning and end of each session, providing participants the opportunity to both affirm their willingness to participate and to consider the content of the recordings before giving final consent. No participants have asked to have recordings partially or fully deleted following completion of any recording session.

It is notable that there are no reasonable expectations of privacy other than for study participants who are hospital patients, and this work was considered to meet the criteria for waiver of informed consent for everyone other than the participants themselves. Regardless, separate media releases were sought from individuals present in the video recordings to use their faces in publications. Some individuals were either unreachable or declined consent to use their faces in print. In those cases, the original faces of those individuals are not shown and their face model representations were rotated by a random matrix that leaves the expressions, motion, and pose unchanged but changes the facial features such that the reconstructed face does not look like the original person.

## Electrode Localization

Electrodes were localized with Brainstorm[49] software using high-resolution postoperative CT scans of participants that were co-registered with preoperative MRI images using FreeSurfer™ [50].

## Data Acquisition

Multimodal behavioral data (egocentric video and eye-tracking) as well as neural activity from 109-218 iEEG contacts (StereoEEG in all cases) were recorded simultaneously during unscripted free viewing sessions in which participants wore eye-tracking glasses while they interacted with friends and family, clinicians, hospital staff, and members of the research team. The timing and duration of recording sessions were determined based on clinical condition, participant preference, and the presence of visitors in the hospital room, when possible.

Behavioral data were captured by fitting each participant with SensoMotoric Instrument's (SMI) ETG 2 Eye Tracking Glasses. An outward facing egocentric world camera recorded video of the scene viewed by participants at a resolution of 1280 x 960 pixels at 24 Hz. Two inward facing eye-tracking cameras recorded eye position at 60 Hz. SMI's iView ETG server application, running on a laptop received and stored streaming data for all modalities from the eye-tracking glasses by way of a USB2.0 wired connection. The iView ETG software also served as an interface to calibrate the eye-tracking glasses to each participant with a three-point calibration procedure that enabled the accurate mapping of eye-tracking data to specific eye gaze locations on video frames, and to initiate and stop the recording of behavioral data.

Electrophysiological activity (field potentials) was recorded from up to 218 iEEG electrodes at a sampling rate of 1 kHz using a Ripple Neuro's Grapevine Neural Interface Processor (NIP).

## Data Synchronization

A MATLAB® script, running on the same laptop as the SMI iView ETG Server software, broadcast numbered triggers every 10 s, injecting them simultaneously into the neural data stream via a Measurement Computing USB-1208FS data acquisition (DAQ) device connected to the NIP's digital port and into the eye-tracking event stream via SMI's iView ETG server application via a sub-millisecond latency local loop back network connection using UDP packets. These triggers were used to align and fuse the heterogeneously sampled data streams after the experiment, as described in the Data Fusion section.

In each session, the recording of neural activity was initiated, followed by simultaneous initiation of recording of eye-tracking and egocentric video via the SMI ETG 2 Eye Tracking Glasses using the SMI iView ETG Software Server. Once the recording of all modalities was underway, the MATLAB® script was initiated to generate and transmit triggers. At the end of each recording session, the tear down sequence followed the reverse order.

## Minimizing Eye-tracking Error and Participant Fatigue

Shift in the position of the eye-tracking glasses was possible if the participant inadvertently touched or moved them during a recording session. Such disruption can introduce systematic error(s) in eye gaze data captured after the disruption(s), although errors can be mitigated with gaze correction (see Data Preprocessing for details). The potential for such an event increases with the duration of a recording session. To minimize the risk of such error(s), we first instructed participants to avoid touching or nudging the glasses during a recording session to avoid disrupting the eye-tracking calibration completed at the beginning of recording sessions. Calibration accuracy was then reverified at the end of recording sessions. Second, we strove to reduce such errors by limiting an individual recording session to 1 hour and including a short break for participants. During this interlude, the recording was terminated, and participants were offered the opportunity to remove the eye tracking glasses before initiation of the next session. The interlude served two purposes: 1) it gave the participant a break from wearing the eye-tracking glasses, helping to alleviate fatigue and discomfort; 2) initiating a new recording allowed the research team to re-secure and re-calibrate the eye-tracking glasses, renewing the accurate mapping of gaze to the egocentric video. Although ≈1 hour recordings were preferred as a best practice, maintaining this practice depended upon participant preference and the number of visitors. In some cases, recording sessions were longer.

**Ergonomic Modifications**

Standard clinical care following iEEG implantation involves the application of a bulky gauze head dressing. This bandaging was applied around the participants' heads to protect the operative sites where the iEEG electrodes were secured with bolts. The dressing also included a chin wrap to provide further support in preventing dislodgement of the iEEG electrodes by securing the connector wires that carry electrical activity to clinical and/or research recording systems. The bandaging typically covered the participants' ears, rendering the temples on the eye-tracking glasses unusable. To overcome this challenge, we modified the structure of the eye-tracking glasses, removing the temples and substituting them with an adjustable elastic band. We attached the elastic band to the frame of the eye-tracking glasses using Velcro patches sown at each end. The modification permitted secure placement of the glasses on the face of a participant, with the elastic band carefully stretched over the head dressing to avoid disturbing the operative sites. To reduce any pressure the eye-tracking glasses placed on the participants' faces as a result of the elastic band alteration, we further modified the glasses by adding strips of adhesive backed craft foam to the nose bridge and upper rims of the frame. These ergonomic solutions enabled correct, robust, and comfortable placement of eye-tracking glasses for each participant with flexibility to adjust to individual bandaging and electrode placement configurations. As an added measure to minimize the possibility of movement for eye-tracking glasses during recording sessions, the USB cable connecting the eye-tracking glasses to the laptop was secured to the participants' hospital gowns near the shoulder with a clip to prevent the weight of the remaining length of cable from pulling on and displacing the glasses during a recording session. Sufficient slack was left in the cable segment between the glasses and the fixation point on the participants' gowns to allow for free head movement while preventing the secured cable segment from pulling on and potentially displacing the eye-tracking glasses.

## Behavioral Experiment

A behavioral psychophysics experiment approved by the University of Pittsburgh's Institutional Review Board was conducted in a cohort of 11 healthy participants (three male, eight female; self reported), whose ages ranged from 19 to 46 years old (mean = 30.54, SD = 6.97).

The behavioral paradigm instructed participants to determine whether the expression on two faces (with the same identity), shown one after the other, were the same or different. A black screen with a fixation cross preceded each trial. Each of the two faces shown in a trial was presented for 500 ms. The two face presentations in each trial were separated by a randomized inter-stimulus interval between 500 ms and 1100 ms long, during which a noise mask was shown on the screen. A randomized inter-trial interval between 500 ms and 1100 ms long separated trials. All the faces presented in this paradigm were generated from a 3D morphable face model with a principal component space in which the unit vectors for canonical expressions (joy, disgust, fear, anger, surprise, sadness) were known. The stimuli varied in expression only, i.e., identity was not varied. A total of 9 expressions were included in the stimulus set. These included joy, sadness, surprise, and 6 expressions that were randomly generated by sampling the face model's expression space. The stimulus set consisted of 10 intensities for each expression and at each intensity 4 modified faces (2 in the direction of a different expression and one each with a higher and lower intensity for the same expression) were generated so they could be paired with the baseline image in a trial. The order in which faces were shown in a trial was randomized, i.e., the baseline facial expression image could be shown before the perturbed version of it or vice versa. This added up to 38 trials for each expression, i.e., 4 trials for each of the 10 expression intensity levels minus the trials which required 1) increasing the intensity for the highest intensity level and 2) decreasing the intensity at the neutral expression. The maximum allowable expression intensity was the same across all expressions, and limited to ensure that presented faces were not exaggerated beyond what a human face normally performs becoming aversive. The paradigm featured sham trials in which the same face was shown twice, which accounted for 20% of the number of normal trials. A single run of the paradigm comprised 410 trials. The overall behavioral experiment consisted of two runs of the paradigm. In the first, the perturbation steps from the baseline face were of the same size in the linear face model space. In the second, they were of the same size in the visual (3D mesh) face space ensuring control over visual differences.

## Data Preprocessing

The behavioral (eye-tracking and video) and physiological (neural) data streams captured during a real world vision recording were preprocessed as follows before data analysis was initiated.

### Eye-Tracking

The eye-tracking data stream was composed of time series data sampled at 60 Hz, where each sample (referred to as an eye tracking trace) contains a recording timestamp, an eye gaze location (X,Y coordinates in the space of egocentric video) and was labeled by the SMI iView ETG platform as belonging to a fixation, a saccade, or a blink. Consecutive eye-tracking traces with the same label (fixation, saccade, or blink) were interpreted as belonging to a single eye-tracking 'event' of that type, whose duration is the difference in recording timestamps of the last and first eye-tracking traces in the block of consecutive traces with the same label (fixation, saccade, or blink). As an example, a set of 60 eye-tracking traces (amounting to 1 s of recorded activity), where the first 30 are labeled as fixation, the next 3 labeled as saccade, followed by the final 27 labeled as fixation, would be interpreted as a fixation event 500 ms long (30 samples at 60 Hz), followed by a saccade event 50 ms long (3 samples at 60 Hz) followed by a fixation event 450 ms (27 samples at 60 Hz). We developed custom Python scripts that parse eye-tracking traces and construct logs of eye-tracking events for each recording session. In addition to the duration of each eye-tracking event, the median gaze location (median was used for robustness to outliers) was logged for each fixation event and the start/end gaze locations were captured for each saccade event. Blink traces also  denoted a loss of eye-tracking (i.e., absence of gaze location) and as a result only the duration of blink events was tracked in the consolidated eye-tracking event logs. Preprocessing of eye-tracking data also incorporated the detection and correction of systematic errors in gaze angle estimation that can be induced by the movement of eye-tracking glasses during recording sessions (e.g., if a participant inadvertently touches and moves the glasses due to fatigue), which disrupts the calibration of eye-tracking glasses (see Data Acquisition for details). Such issues were detected by manually viewing all experimental recordings using SMI's BeGaze application, which renders eye-gaze, audio, and egocentric video together. The disruption of calibration for eye gaze tracking was visually detectable when viewing egocentric video overlaid with eye-tracking and audio because visual behavior was altered such that the gaze data fails to make sense consistently after loss of eye-gaze calibration (e.g., the subject was scrolling through a phone or reading a book or watching tv or talking to someone, but the gaze location was visibly shifted away from the obvious target). These issues were corrected using the SMI BeGaze application, which allows for the application of a manual correction to eye gaze at any time point in a recording, which applies to all eye gaze data following the corrected time point. The corrections were verified by reviewing the video that followed the correction, to ensure that corrected eye gaze data made sense consistently. Corrections to eye-tracking data preceded preprocessing in such cases.

**Intracranial Recordings**

Fixation Related Potentials (FRPs) and Fixation Related Broadband High-Frequency Activity (FBHA) were extracted from the raw iEEG recordings for statistical analysis using MATLAB®. FRPs were extracted using a fourth-order Butterworth bandpass ([0.2 Hz, 115 Hz]) filter to remove slow linear drift and high-frequency noise, followed by line noise removal using a fourth-order Butterworth band stop ([55 Hz, 65 Hz]) filter, and aligning the resulting response potential with eye-tracking data to obtain FRPs. FBHA extraction involved two steps. First, the raw signal was filtered using a fourth-order Butterworth bandpass ([1 Hz, 200 Hz]) filter followed by line noise removal using notch filters at 60, 120, and 180 Hz to obtain local field potentials. Next, power spectrum density (PSD) between 70 and 150 Hz was calculated for the local field potentials with a bin size of 2 Hz and a time-step size of 10 ms using Hann tapers. For each electrode, the average PSD across the entire recording was used to estimate a baseline mean and variance of the PSD for each frequency bin. The PSD was then z-scored using these baseline measurements for each frequency bin at each electrode. BHA was estimated by averaging the z-scored PSD across all frequency bins (excluding the line noise frequency bin at 120 Hz). Finally, FBHA was obtained by aligning the Broadband High-Frequency Activity with eye-tracking data to obtain FBHA.

iEEG recordings were subjected to several criteria for inclusion in the study. Any recordings with ictal (seizure) events were not included in the study. Artifact rejection heuristics were implemented to avoid potential distortion of statistical analyses due to active interictal (between seizure) or outliers. Specifically, we evaluated the filtered iEEG data against three criteria that were applied to each sample, i.e., each time point in iEEG recordings, which corresponds to 1 ms of neural activity. These criteria were applied to the filtered iEEG signal for each electrode, as well as the averaged (across all electrodes) iEEG signal. The first criterion labels a sample as 'bad' if it exceeds 350 μV in amplitude. The second criterion labels a sample as bad if the maximum amplitude exceeds 5 standard deviations above/below the mean. The third criterion labels a sample as bad if consecutive samples (1 ms apart at a 1000 Hz sampling rate) change by 25 μV or more. For the averaged iEEG signal, any sample satisfying any of these three rejection criteria is labeled as bad. Further, if more than ten electrode contacts satisfy the bad sample criterion for a particular sample, it is labeled as a bad sample. Less than 10% of the samples in experimental recordings were labeled as bad samples. All data types were dropped from analysis for fixations that contained bad samples in either FRPs or FBHA.

**Face Detection**

Egocentric world camera video recordings included a range of visual stimuli present in the room, including objects, people, and faces. We processed these recordings to detect faces, the primary object of interest in this study. Deep Learning based computer vision models[51,52] developed for large-scale face detection and recognition applications were used to detect faces present in each video frame of egocentric video recordings. Manual review of egocentric videos with bounding boxes that annotated identifying detected faces showed model performance was robust to the variability of conditions present in the egocentric video recordings. Failure to detect a face was rare, usually involving heavy occlusion, extremely poor lighting, or both. Cumulatively, 761,510 faces were detected across 1,136,208 frames of video corresponding to 11 hours of recordings, which required ≈40 hours on a single NVIDIA® GeForce GTX 1080TI GPU. In addition to face detection, these models also generate a 512-dimensional embedding in a face space that could be used to train classifiers to identify different individuals present in the video recordings.

**Face Identification**

A neural network was trained to perform identity classification on faces detected across all video recordings for each participant. The network architecture featured two densely connected layers (128 ReLU units each) that were subjected to a 50% dropout rate to avoid overfitting. 512-dimensional embeddings generated by face detection models were used to predict identity. To prepare data for model training, identity labels were assigned manually to all faces present in a subset of video frames which corresponded to the beginning of each fixation. The annotation typically required 2 hours of effort to annotate a 1 hour video recording. Depending upon the participants' activities during a recording session, faces could also be detected on television screens, mobile phone screens, magazines, and even arise from false positives (<1%), none of which were in scope for this study. These extraneous faces arose in training data and were given a catch-all identity label ("other") that separated them reliably from the faces of real people in the room. Models were trained using 5-fold cross-validation and high accuracy (<0.1% misclassification) and class balanced accuracy (since people were present in the recording for different amounts of time) were observed on held out data.

The trained identity classification networks for each patient were then used to label all the faces detected in each video frame of all their recording sessions. A final manual review of the fully annotated video was performed to ensure undesirable and unforeseen issues did not arise, e.g., mislabeling sparsely present individuals as "other."

## Face Parameterization

Each face detected in egocentric video recordings was represented in a linear face model that represented a 3D structure for them[11,12]. Recent advances have enabled robust estimation and high-fidelity reconstruction of 3D faces from monocular images that capture a 2D view of the original face. Face AI models perform reconstructions estimating the head pose, shape, texture, expression, and eye gaze of faces present in 2D images while accounting for extraneous factors, such as camera position and lighting parameters, in a way that aligns important facial landmarks, facial appearance, and minimizes pixel level loss for the reconstructed face compared against the original 2D face image.

Here, we parameterized faces in each egocentric video frame to obtain the head pose ($\theta \in R^3$), as well as their shape ($s \in R^{80}$), texture ($t \in R^{80}$), and expression ($e \in R^{64}$) in a 3D face model defined by a principal component space (an ortho-basis for faces)[11,12] using Deep 3D Face[10]. Separately, we obtained estimates for eye gaze ($g \in R^2$) for faces in each frame of egocentric video recordings using a state of the art neural network[13]. Combining these results in a 229-dimensional representation for each face in each video frame of the egocentric video recordings. While the findings presented here result from the specific operationalization of facial features by the Deep 3D Face AI model, we observed comparable results with other Face AI models[53] that provide high dimensional representations of faces present in the egocentric video.

## Face Dynamics Model

A linear dynamical systems model was used to identify a low dimensional latent space where the trajectories representing the dynamics of parameterized faces could be embedded and recovered to reconstruct the original 229-dimensional representation reliably. Head pose and eye-gaze were low dimensional variables whose dynamics are tracked separately, i.e. they are not embedded in the state space. Thus, the state space model represents $\mathbb{R}^{224}$ dimensional inputs spanning shape (s), texture (t), and expression (e) in a latent space ($x \in \mathbb{R}^{30}$). The model structure is described below and follows smooth linear dynamics ($A$), linear coordinate transformations into the latent space ($B$), and a linear read out ($C$) of the latent variables back into the original $\mathbb{R}^{224}$ face representation.

$$x_{t+1} = Ax_t + B \begin{bmatrix} s \\ t \\ e \end{bmatrix} \quad \text{Equation 1}$$

$$\begin{bmatrix} \hat{s} \\ \hat{t} \\ \hat{e} \end{bmatrix} = Cx \quad \text{Equation 2}$$

For each participant, these models were trained and validated on face trajectories from un-fixated faces that were not used in analysis. Validation also included a qualitative component, where face videos reconstructed from latent representations for held out data were reviewed alongside the original faces which were not embedded in the model. The models were used to generate trajectory embeddings for fixated faces that were used in analysis against brain activity.

**Resting Face Estimation**

The resting facial expression for each individual present in egocentric video recordings was estimated using parameterized representations of un-fixated faces that were not used in analysis. The resting facial expressions were then subtracted from all fixated faces that were used in analysis against brain activity.

The first step in estimating resting facial expression was to regress out the effects of head pose on face parameters. This was done by training a multiple regression model to predict the value of face shape, texture and expression parameters based on head pose. Removing values predicted by the regression model provided a head pose corrected parameterization for each fixated face. Subsequently, the average head pose corrected shape, texture, and expression for each identity were computed as the resting facial expression.

## Data Analysis

Precise alignment of the heterogeneous behavioral (eye-tracking), environmental (egocentric video) and physiological (neural) data streams was essential for robust analysis. This was achieved using eye-tracking as a reference modality against which video and intracranial recordings were aligned in time, as described previously[45]. All analyses in this study were anchored to behavioral events (fixations) and each fixation was mapped to corresponding egocentric video frames, as well as brain activity after the fixation began, i.e., fixation locked brain activity that was comprised of FRP and FBHA. Analysis combined FRP and FBHA in an information centric approach as both signals have been shown to contain complementary and independent information relevant to cognition[54,55].

### Data Preparation

Fixations were determined to be on a face if the eye-gaze at the beginning of a fixation was on a person's face in the corresponding egocentric video frame. This was operationalized by determining if eye-gaze coordinates fell within a face bounding box identified by computer vision models[51,52]. Face fixations were filtered to ensure that they were 300ms or longer. Next, they were filtered to ensure that no fixation contained brain activity that has been characterized as having bad samples. Next, face parameters for the fixated person were retrieved for all the egocentric video frames corresponding to the face fixation, and fixations where this was not possible for any reason were filtered out (e.g., the face was not detected due to being occluded by obstacles like another person crossing them). While the identity and expressions of people with varying familiarity to the research participant, e.g., spouses, family, friends, researchers, etc. were modeled in this study, the influence of familiarity on the neural encoding of expressions is a rich area for future work.

The fixations that satisfied these data quality criteria were then assembled into a dataset (X,Y), where $X \in R^p$ represented face trajectories in the face dynamics model and $Y \in R^q$ represented fixation locked brain activity (FRP and FBHA) for all intracranial electrodes implanted in a participant's brain. The dimensionality p of the face trajectories corresponded to collecting the pose ($\theta \in R^3$), eye gaze ($g \in R^2$), and latent face trajectories $x \in R^{30}$ for 7 frames (corresponding to ≈300 ms), which resulted in $p = 7 \times (3 + 2 + 30) = 245$ for basic face reconstruction and $p = 7 \times (2 + 30) = 224$ for the dataset to study facial expressions as deviations from the resting face. The dimensionality q of brain activity depended upon the number of electrodes, which differed for each participant. For 300 ms, FRPs sampled at 1 kHz corresponded to 300 dimensions for each electrode and FBHA sampled at 100 Hz corresponded to 30 dimensions for each electrode, which resulted in $q = E \times (300+30)$. Participants in this study were implanted with anywhere between 109 to 218 electrodes which corresponded to $q \in R^{35970 \text{ to} 71940}$. The number of fixations ranged from $N \approx 10^2$ to $10^3$ across different participants. For computational modeling, the fixations are then split into training (80%) and test (20%) sets that are rotated for 5-fold cross validation which ensures that every fixation is in the test set once.

## Computational Model

Canonical correlation analysis[56] seeks to model the covariability between two multivariate datasets ($X \in R^p$, $Y \in R^q$) as a small number of strongly correlated latent variables (Canonical Components), to understand the relationship between them. In contemporary machine learning terms, it is also described as a type of latent multi-view representational learning approach. In low dimensional data rich settings where $N > p, q$, CCA can be implemented using a Singular Value Decomposition (SVD) on $\Sigma_{YY}^{-\frac{1}{2}}\Sigma_{YX}\Sigma_{XX}^{-\frac{1}{2}}$ ). However, this approach does not scale to high dimensions where $N \ll p, q$ due to challenges with inverting $\Sigma_{XX}$, $\Sigma_{YY}$. Different approaches have been proposed to address these challenges with high dimensional data[57–61] including the idea of regularized models to estimate sparse canonical vectors[59,60] .

Here, Sparse CCA was implemented by an iterative penalized least squares algorithm[60] which used regularized regressions in an alternating manner to estimate canonical vectors for (X and Y) one canonical component at a time. Given a centered dataset $X \in R^{N \times p}$, $Y \in R^{N \times q}$ with sample covariances, $\hat{\Sigma}_{XX} = \frac{1}{N}X^TX, \hat{\Sigma}_{YY} = \frac{1}{N}Y^TY, \hat{\Sigma}_{YX} = \frac{1}{N}Y^TX$, where the first k−1 pairs of canonical vectors $(\hat{w}^l_{face}, \hat{w}^l_{brain})$ ∀ l ∈ (1,–k − 1) have been estimated, the k[th] canonical vectors $(\hat{w}^k_{face}, \hat{w}^k_{brain})$ are estimated by solving

$$
\begin{aligned}
(\hat{w}^k_{brain}, \hat{w}^k_{face}) = \underset{w^k_{brain}, w^k_{face}}{\operatorname{argmin}} \; & \frac{1}{2N}\sum_{i=1}^{N}\left(Y_i^T w^k_{brain} - X_i^T w^k_{face}\right)^2 + \\
& w^k_{brain}\left(\sum_{l<k}\hat{\rho}_l \hat{\Sigma}_{YY}\hat{w}^l_{brain}\hat{w}^l_{face}\hat{\Sigma}_{XX}\right) + \\
& P_Y(w^k_{brain}) + P_X(w^k_{face}) \\
& s.t.\; w^k_{brain}\hat{\Sigma}_{YY}w^k_{brain} = 1,\; w^k_{face}\hat{\Sigma}_{XX}w^k_{face} = 1
\end{aligned}
$$

Equation 3

where $P_Y(w^k_{brain})$ and $P_X(w^k_{face})$ are regularization functions that may reflect the type of penalization in effect such as group lasso, trend filtering, etc. Here, we chose elastic penalties (Equation 4) that combined sparse feature selection with a smooth distribution of weights over the selected features. It is notable that although the optimization problem is nonlinear in nature, the model structure itself is linear.

$$
P(w, \lambda, \alpha) = \lambda\left(\alpha||w||_1 + \frac{(1-\alpha)}{2}||w||_2^2\right)
$$

Equation 4

The optimization problem in Equation 3 was solved using the algorithm described in[60].

**Training**

Training data were standardized (demeaned and scaled to unit variance) prior to model training. Models were trained with 5-fold cross-validation, which allowed each sample (fixation) to be in the test set once. The use of an elastic penalty function required choosing two parameters optimally for each canonical component for each of the 5 models (one per fold) trained in this way. The first was the regularization penalty ($\lambda_{brain}$, $\lambda_{face}$) and the second is the elastic penalty ($\alpha_{brain}$, $\alpha_{face}$). Both were identified during an additional 5-fold cross-validation procedure within the training data, i.e., an additional inner cross-validation loop. A distinct relationship between $\alpha$ and $\lambda$ (for brain activity and facial features) emerged during hyperparameter selection, where the amount of L1 penalty they collectively enforced ($\alpha \times \lambda$) remained identical, i.e., increasing $\alpha$ led to a lower optimal $\lambda$ and decreasing $\alpha$ led to a higher optimal $\lambda$ such that their product, i.e., the L1 penalty, remained nearly constant. This observation was interpreted as a property of the data rather than the algorithm and was used to optimize the model training procedure by choosing a low value of $\alpha = 0.1$, which was fixed and $\lambda$ was the only hyper parameter being optimized. The choice of $\alpha$ ensured the L2 penalty term, weighed as $\frac{1}{2}(1-\alpha)$, provided greater smoothing across the aspects of brain activity and facial features that were selected by the model.

In terms of model selection, canonical components which exhibited statistically significant correlation during the inner cross-validation loop were preserved, and those that did not had their weights zeroed out. Models estimated up to 20 canonical components during training but the number of canonical vector pairs that survived cross-validation did not exceed 10. All k pairs of canonical vectors that survived cross-validation were arranged into two matrices $\widehat{W}_{face}^{K} \in \mathrm{R}^{k \times p}$ and $\widehat{W}_{brain}^{K} \in \mathrm{R}^{k \times q}$. The approach of estimating a larger model ensured that no useful relationships were missed. The canonical space is also referred to as the neuro-perceptual space here since it is jointly learned from brain activity and face trajectories.

## Inference

Held out test data were centered according to the mean and variances estimated from training data prior to inference. Face trajectories and brain activity for these fixations were first projected into the neuro-perceptual space ($Y_i^{CC} = \widehat{W}_{brain}^{K} \times Y_i$ and $X_i^{CC} = \widehat{W}_{face}^{K} \times X_i$). Top level statistics in the form of d' were then computed from these neuro-perceptual representations to assess model performance. Specifically, pairwise classification accuracy was computed for each held out fixation by comparing the distance between its face trajectory ($X_i^{CC}$) and brain activity ($Y_i^{CC}$) against distances with all other held out fixations ($j \neq i$) in the neuro-perceptual space. Distances were weighted by singular values D obtained from SVD during initialization of the canonical vectors, as described in[60]. Similar classification accuracies were computed for comparisons between fixations on different individuals (person i ≠ person j) and for different fixations on the same person (j≠i and person j = person i). In this way, classification accuracy was calculated for each fixation. The resulting distribution of classification accuracies for all fixations is summarized by its median, which best represents the central tendency even when the distribution is asymmetric, as in the case of accuracies.

$$\text{Accuracy} = \text{median}\left(\frac{1}{N-1}\sum_{\substack{j \\ j\neq i}}^{N} \mathbb{I}_{||D(X_i^{CC}-Y_i^{CC})||_2<||D(X_i^{CC}-Y_j^{CC})||_2}\right)$$

Equation 5

The classification accuracy can be estimated by comparing the neuro-perceptual face representation of a fixation with the neuro-perceptual brain activity representation of all other fixations (described in Eq. 5) or by comparing the neuro-perceptual brain representation of a fixation with the neuro-perceptual face representation of all other fixations, which changes Equation 5 to use $\mathbb{I}_{||D(X_i^{CC}-Y_i^{CC})||_2<||D(X_j^{CC}-Y_i^{CC})||_2}$ instead. In practice, both resulted in similar top-level statistics. Finally, the median accuracy is transformed into d' as per $\mathrm{d}' = \Phi^{-1}(\text{Accuracy}) - \Phi^{-1}(1-\text{Accuracy})$, where Φ is the CDF of the standard normal distribution. Note that d' is an effect size measure on a similar scale as Cohen's d and thus values of 0.8, 1.3, and 2.0 are generally considered "large", "very large", and "huge" effect sizes[62].

***Reconstructed Faces***

Faces for each fixation were predicted from corresponding fixation locked brain activity by projecting brain activity into the neuro-perceptual space ($Y_i^{CC}$) and inverting that projection to obtain a predicted face trajectory ($\hat{X}_i = Y_i^{CC} \times \hat{W}_{face}^{K}{}^{\dagger}$) , where † represents the pseudoinverse of a sparse low rank projection matrix. The predicted face trajectory was read out from the state space of the face dynamics model and into the linear face model's coefficient space, where it was rendered into a series of face images that were compiled into videos. The face model lacked the ability to visualize eye-gaze, i.e., rotate the eyeballs of a visualized face according to the original (or neurally predicted) eye-gaze for rendering purposes. This gap was addressed with a procedure that rotated the eyeballs of faces in the 3D face mesh space before rendering them into a 2D image. The method provided an approximate rendering of the estimated (or neurally predicted) eye-gaze for fixated faces that was used purely for visualization purposes in Fig. 2A, and Movie 2; https://tinyurl.com/4D6F76696532.

This visualization procedure was also used to visualize how moving along different dimensions of the neuro-perceptual space affected face appearance and facial motion, i.e., aspects of neural tuning can be revealed as shown in Fig. 3B, Movie 5; https://tinyurl.com/4D6F76696535.

***Reconstructed Neurodynamics***

Brain activity for each fixation was predicted from the fixated face by projecting the face trajectory for it into the neuro-perceptual space ($X_i^{CC}$), and inverting that projection into the space of brain activity ($\hat{Y}_i = X_i^{CC} \times \hat{W}_{brain}^{K\,\dagger}$). Predicted brain activity was compared with original neural activity for all electrodes at all time points for both FRPs and FBHA. Aspects of predicted brain activity that were significantly correlated with the original were identified using permutation tests.

The neurodynamics of reconstructed brain activity were visualized with brain heatmaps obtained using a three-dimensional gaussian kernel (bandwidth = 15mm) to smooth (statistically significant) correlations for each electrode on a common brain[63,64], at each time point (no temporal smoothing was performed). In this way, a movie depicting the strength of reconstructed neurodynamics at each time point over the course of a fixation was obtained (Movie 3; https://tinyurl.com/4D6F76696533).

This visualization procedure was also used to visualize how moving along different dimensions of the neuro-perceptual space affects neurodynamics, i.e., aspects of neural tuning can be revealed as shown in Fig. 3B, Movie 5; https://tinyurl.com/4D6F76696535.

**Probing Population Tuning in the Neuro-Perceptual Space**

A step in the model's neuro-perceptual space changed the predicted face trajectory and the predicted pattern of brain activity. The step sizes of these changes were linearly dependent on the step size in the neuro-perceptual space. This coupling enabled testing hypotheses about neural population tuning by exploring the relationship between the geometry of the data manifold of brain activity and of stimuli (face trajectories). Here, this was done by relating the Euclidean distances between the face ($Y_i^{CC}$) and neural representation ($X_i^{CC}$) of each fixation, i.e., the error of neural prediction, to the intensity of expression (size of deviation away from the norm face).

**Permutation Tests**

Three types of permutation tests were implemented to estimate the statistical significance thresholds for top level statistics (d') and reconstructed brain activity. Permutation testing (1000 permutations) was implemented for each participant, for both the basic face reconstruction models and for the norm-based models. In the first type of permutation test, the pairing of brain activity and facial features were permuted for samples, i.e., the brain activity associated with a face fixation was permuted and assigned to a different face fixation. In the second type of permutation tests, fixation locked brain activity from fixations that were on objects other than faces was paired with facial features. In the third type of permutation test, randomly selected facial features were paired with randomly sampled brain activity that was not anchored to fixations at all, instead of the original fixation locked brain activity associated with each face fixation.

Statistical significance for d' was calculated from a null distribution of those statistics estimated from 1000 permutations. Statistical significance for correlations between actual and reconstructed brain activity were determined from a null distribution of those correlations estimated from 1000 permutations.

Finally, face predictions from the second type of permutation tests were also rendered to qualitatively assess what happens to the predicted faces when spurious brain activity was injected into the model. Face predictions from permutation tests were observed to be very close to the origin of the model with expressions and motions that resembled noise around the origin, i.e., the mean face of the model. Geometrically, this suggested that non-face fixation locked brain activity disappeared into the null space of the model. Although the model produced 'a' face, it was unavoidable for it to do so, because it was a face model.

**Group Analysis**

Linear mixed effects models were implemented for group level analysis of reconstructed neurodynamics[65]. Since the location of intracranial electrodes differed for each participant, the first step of the analysis was to project the original and model predicted neurodynamics (or the neural prediction error) for each participant onto a common cortical surface[63,64] with a gaussian kernel (bandwidth = 15 mm), like what was done to visualize reconstructed neurodynamics. The original and model predicted neurodynamics obtained for each vertex of the common cortical surface were then combined to fit linear mixed effects models for vertices where brain activity from two or more participants was observed. To account for multiple comparisons across all cortical vertices and over time, the statistical significance of model fits was FDR corrected[66] ($p<0.05$). Finally, statistically significant model fits were visualized on the brain.

A similar approach was taken for group level analysis for the neurodynamic correlates of Weber's law for facial expressions. In this instance, the error of neural prediction was calculated for each cortical vertex for each fixation for each participant and linear mixed effects models were trained to predict it using facial expression intensity.

**Correlation maps between brain activity in controlled experiments and natural recordings**

Correlation maps between brain activity in a controlled ("localizer") experiment and during real world behavior were computed for each participant. The localizer consisted of participants viewing static images of faces, houses, bodies, words, hammers, and phase scrambled face on a computer screen while performing a 1-back task[54,55]. Due to many differences in task and stimuli, the localizer was used just to determine if face activation maps were similar between real world faces and faces in a controlled experiment. Correlation maps were estimated by first computing the area under the curve for each electrode for response potentials and for broadband during the first 300 ms (the fixation duration used in analysis) of brain activity in response to face stimuli. The areas under the curve for all electrodes were correlated using Spearman's correlation, separately for response potentials and broadband. The two correlation values were then averaged using Fischer's method[67] to obtain a correlation value that reflected the shared variance between brain activity during a controlled experiment and during natural behavior in the real world.

### Analysis of Behavioral Data

Participant responses were tallied against ground truth to determine response accuracy/error as a percentage, after removing sham trials. This amounted to counting the number of trials where the participant response rated faces as having different expressions and dividing them by the total number of trials in consideration. Before being subject to this basic arithmetic, the trials were partitioned based on the intensity of expressions and the accuracy of behavioral responses in those trials were tracked as a function of expression intensity, i.e., distance from the neutral face.

## Code Availability

Code for neuro-perceptual spaces: https://github.com/arishalreja/real_world_face_perception

## Data Availability

Data are available upon reasonable request and IRB approval.

## Acknowledgements

We thank the patients for participating in the iEEG experiments; the UPMC Epilepsy Center monitoring unit staff, physicians and administration, particularly A. Bagic, C. Plummer, M. Millen, and D. Crawford, for their assistance and cooperation with our research; K. Rupp, S. Ramakrishnapillai, J. Hect, M. Richey, and E. Harford for their help with data collection; Q. Ma, N. Silverling and T. Gautreaux for their help with annotations of the egocentric video; R. Knight, R. Kass, C. Becker, W. Lipski, D. Geng, M. Zhou, I. Yu, and M. Wang for their helpful comments on the manuscript. This work was supported by grants from the National Institutes of Health (R01MH132225, R01MH107797, P50MH109429), the National Science Foundation (1734907), the Richard King Mellon Foundation, and the Beckwith Foundation.

## Author Contributions

Conceptualization: AA, ASG, LPM, LSP, MJW, RMR, TJA

Methodology: AA, ASG

Investigation: AA, ASG, MJW, RMR, TJA

Visualization: AA, ASG, MJW

Funding acquisition: ASG

Project administration: ASG

Supervision: ASG

Writing – original draft: AA, ASG

Writing – review & editing: AA, MJW, RMR, LPM, LSP, TJA, ASG

## Competing Interests

AA and ASG are inventors on a pending patent application (PCT/US25/019965) filed by the University of Pittsburgh that is related to this work.

## Materials and Correspondence

Correspondence and Materials Requests should be addressed to Arish Alreja (arish.alreja@pitt.edu) and Avniel Singh Ghuman (ghumana@upmc.edu).

# Supplementary Information:
# Neural encoding of real world face perception

**Authors:** Arish Alreja[1,2,3,4,*], Michael J. Ward[4,5], Lisa S. Parker[6], R. Mark Richardson[5,7,8], Louis-Philippe Morency[9], Taylor J. Abel[2,3,4,10,11], Avniel Singh Ghuman[2,3,4,10]

**Affiliations:**

[1]Machine Learning Department, Carnegie Mellon University, Pittsburgh, USA
[2]Neuroscience Institute, Carnegie Mellon University, Pittsburgh, USA
[3]Center for the Neural Basis of Cognition, Carnegie Mellon University and University of Pittsburgh, Pittsburgh, USA
[4]Department of Neurological Surgery, University of Pittsburgh, Pittsburgh, USA
[5]David Geffen School of Medicine, University of California Los Angeles, Los Angeles, USA
[6]Center for Bioethics & Law, School of Public Health, University of Pittsburgh, Pittsburgh, USA
[7]Department of Neurosurgery, Massachusetts General Hospital, Boston, USA
[8]Harvard Medical School, Boston, USA
[9]Language Technologies Institute, Carnegie Mellon University, Pittsburgh, USA
[10]Brain Institute, University of Pittsburgh, Pittsburgh, USA
[11]Department of Bioengineering, University of Pittsburgh, Pittsburgh, USA

## Supplementary Results

### Cortical areas with significantly reconstructed brain activity

The areas of cortex where electrodes were implanted to record brain activity differed for each participant based on clinical considerations. Brain activity from electrodes in multiple brain areas, including traditional face areas in ventral temporal cortex, was reconstructed significantly ($p<0.01$; permutation tests) for each participant, both in sample (Fig. 3A and Movie 3; https://tinyurl.com/4D6F76696533) and out of sample (Movie 3; https://tinyurl.com/4D6F76696533). Notably, electrodes located in areas corresponding to the putative social-vision pathway showed the most robust reconstructions (Table S1).

| Cortical Area (Significant/Total Electrodes) | Participant (Number of Face Fixations) | | | | | |
|---|---|---|---|---|---|---|
| | 1 (1300) | 2 (307) | 3 (2283) | 4 (618) | 5 (207) | All |
| Amygdala | 0/0 | 1/1 | 0/0 | 0/0 | 0/0 | **1/1** |
| Bank of Superior Temporal Sulcus | 1/4 | 0/0 | 1/1 | 0/0 | 0/0 | **2/5** |
| Caudal Anterior Cingulate | 0/4 | 0/0 | 0/0 | 0/3 | 0/0 | **0/7** |

| Caudal Middle Frontal | 0/0 | 0/0 | 0/0 | 7/36 | 0/0 | **7/36** |
|---|---|---|---|---|---|---|
| Cuneus | 0/2 | 7/7 | 1/1 | 0/0 | 4/5 | **12/15** |
| Entorhinal | 0/0 | 0/0 | 0/0 | 0/1 | 0/0 | **0/1** |
| Fusiform | 1/3 | 0/6 | 7/8 | 0/0 | 0/1 | **8/18** |
| Hippocampus | 0/0 | 1/5 | 0/0 | 2/10 | 0/0 | **3/15** |
| Inferior Parietal | 8/10 | 8/12 | 5/8 | 0/0 | 3/16 | **24/46** |
| Inferior Temporal | 1/2 | 1/5 | 6/8 | 0/1 | 1/6 | **9/22** |
| Insula | 1/7 | 0/1 | 4/4 | 2/5 | 0/3 | **7/20** |
| Isthmus Cingulate | 1/1 | 2/2 | 3/4 | 0/0 | 0/0 | **6/7** |
| Lateral Occipital | 3/4 | 8/13 | 4/4 | 0/0 | 3/6 | **18/27** |
| Lateral Orbitofrontal | 0/3 | 0/0 | 0/0 | 1/8 | 0/0 | **1/11** |
| Lingual | 0/3 | 1/2 | 7/7 | 0/0 | 3/6 | **11/18** |
| Medial Orbitofrontal | 0/0 | 0/0 | 0/0 | 0/2 | 0/0 | **0/2** |
| Middle Temporal | 2/4 | 3/8 | 2/5 | 5/15 | 0/2 | **12/34** |
| Paracentral | 0/0 | 0/1 | 7/8 | 0/0 | 2/5 | **9/14** |
| Parahippocampal | 0/0 | 0/2 | 6/12 | 0/0 | 0/0 | **6/14** |
| Pars Opercularis | 1/4 | 0/0 | 0/0 | 4/16 | 0/0 | **5/20** |
| Pars Orbitalis | 0/0 | 0/0 | 0/0 | 0/1 | 0/0 | **0/1** |
| Pars Triangularis | 1/1 | 0/0 | 0/0 | 1/6 | 0/0 | **2/7** |
| Pericalcarine | 4/7 | 4/4 | 7/7 | 0/0 | 0/4 | **15/22** |
| Post Central | 0/4 | 0/0 | 3/3 | 0/0 | 4/14 | **7/21** |
| Posterior Cingulate | 0/0 | 0/0 | 4/4 | 0/3 | 0/2 | **4/9** |
| Pre-Central | 1/1 | 0/0 | 0/0 | 7/34 | 2/9 | **10/44** |
| Precuneus | 4/6 | 6/14 | 14/16 | 0/0 | 1/12 | **25/48** |
| Rostral Anterior Cingulate | 1/1 | 0/0 | 0/0 | 2/3 | 0/0 | **3/4** |
| Rostral Middle Frontal | 3/6 | 0/0 | 0/0 | 1/22 | 0/0 | **4/28** |
| Superior Frontal | 1/3 | 0/0 | 0/0 | 8/51 | 1/3 | **10/57** |
| Superior Parietal | 0/0 | 4/6 | 9/9 | 0/0 | 3/10 | **17/25** |
| Superior Temporal | 4/8 | 3/7 | 2/2 | 0/0 | 1/2 | **10/19** |
| Supramarginal | 9/22 | 7/11 | 6/8 | 0/0 | 0/13 | **22/54** |
| Temporal Pole | 0/0 | 0/0 | 3/3 | 0/0 | 0/0 | **3/3** |
| Transverse Temporal | 1/1 | 1/2 | 2/2 | 0/0 | 1/7 | **5/12** |
| **Total Electrodes** | **47/110** | **57/109** | **103/124** | **40/218** | **29/126** | **276/687** |

**Table S1.** Distribution of electrodes in different cortical areas per the Desikan-Kiliany Atlas[1], that were significantly reconstructed ($p<0.01$; permutation tests) based on out of sample reconstructions across all participants.

## Reconstruction of Prototype Anchored Face Representations

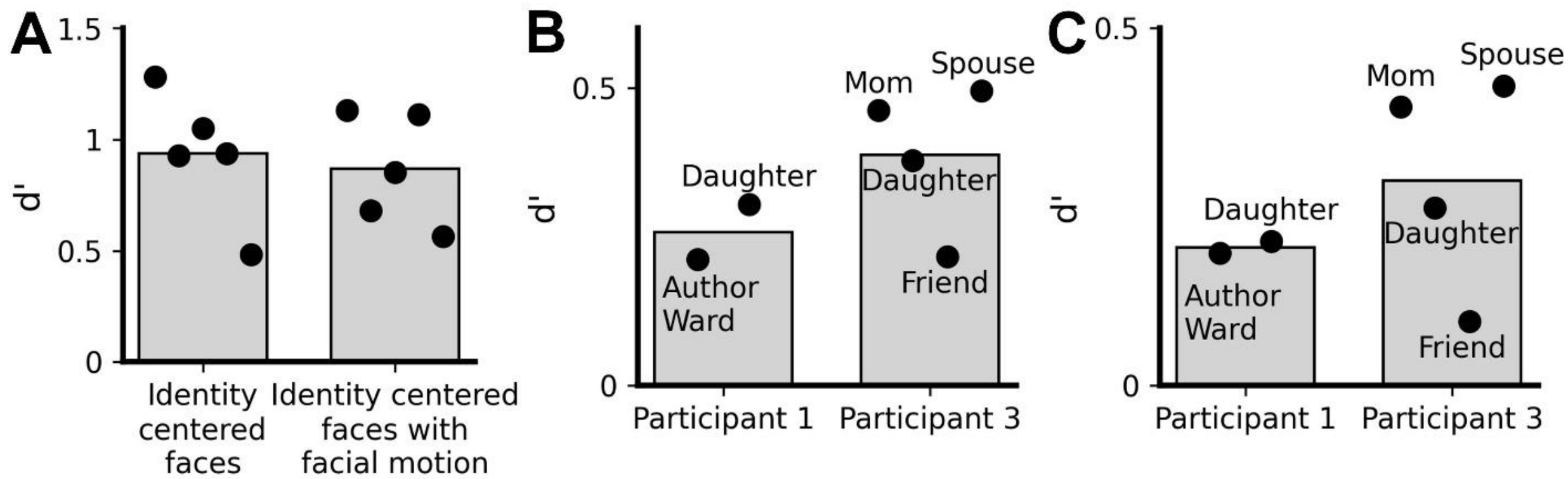

**Fig. S1. Reconstruction Accuracy (d')** for **(A)** Identity centered faces was significant with and without facial motion (p<0.05 for 4 participants, p<0.10 for Participant 5; using permutation tests). **(B)** Cross identity reconstruction accuracy for faces of individuals excluded from training data was significant without facial motion (p<0.05; permutation tests) **(C)** and after incorporating facial motion (p<0.05 – for all except for Participant 3's friend; using permutation tests).

Centered face representations were modeled for all five participants after removing the resting facial appearance of each individual they fixated upon. Reconstruction accuracy (d') was significant for held-out fixations (Fig. S1A). Additionally, cross identity facial reconstruction, where a model was trained on data with person A and person B's centered facial expressions to predict person C's centered facial expressions was undertaken for two participants with enough data i.e., at least one individual with a minimum of 200 fixations on them. Significant reconstruction accuracy (d') was observed (Fig. S1B) for cross-identity decoding of facial expressions showing out of identity generalization of facial expression prediction. Notably, these results remained consistent even when facial motion was incorporated into the models (Fig. S1A and Fig. S1C).

## Weber’s Law for Facial Expression

The neural prediction error reflects a lower (neural) bound on the discriminability of facial expressions in sampled areas of cortex. Trends of neural prediction error vs. expression intensity (Fig. 4A) were quantified using a linear parameterization of Weber’s law[2] (Table S2) whose relative simplicity explained the data (adjusted $r^2$ across participants = 0.42±0.18) as well as an exponential parameterization[3] (adjusted $r^2$ across participants = 0.43±0.18).

| Participant | Slope | | Intercept | |
|---|---|---|---|---|
| | **Estimate** | **95% CI** | **Estimate** | **95% CI** |
| 1 | 0.6426 | [0.486, 0.7992] | 0.0259 | [-0.0282,0.0799] |
| 2 | 0.1963 | [0.1511,0.2415] | 0.02615 | [0.0025,0.0497] |
| 3 | 0.1747 | [0.1082,0.2411] | 0.1947 | [0.1601,0.2292] |
| 4 | 0.6563 | [0.2783,1.034] | 0.0721 | [-0.0789,0.2232] |
| 5 | 0.2818 | [0.2154,0.3482] | 0.2974 | [-0.0015,0.0508] |

Table **S2.** Estimated linear fit parameters for Weber’s law in the neuro-perceptual space.

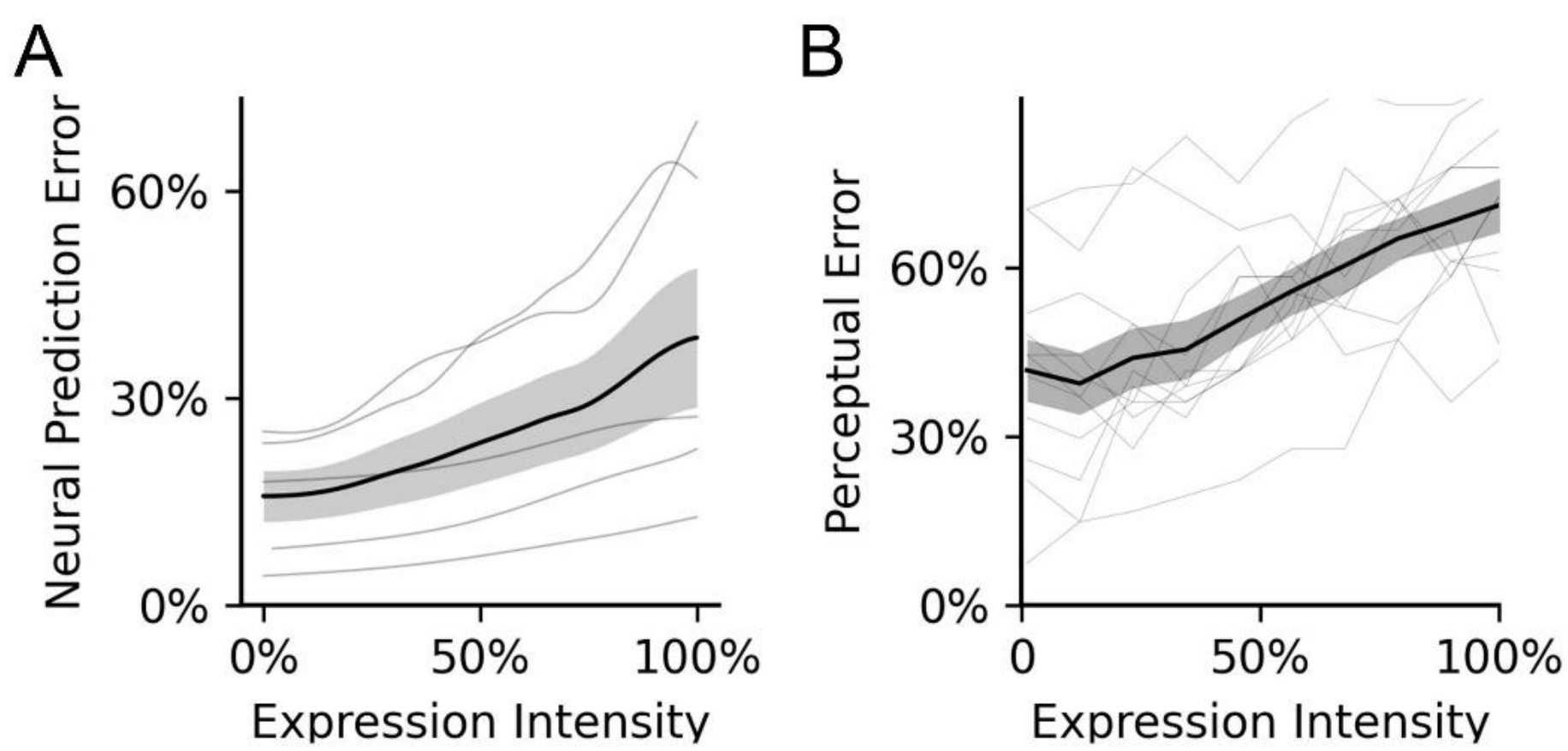

**Fig. S2. (A)** Neural prediction errors for facial expression increased with the intensity of expressions in the 3D face mesh space (correspondeding to low level, pixelwise visual differences in faces), consistent with the face model space (Fig. 4A). **(B)** Behavioral responses showing decreasing sensitivity for facial expressions as intensity of expression increased in the 3D face mesh space, consistent with the face model space (Fig. 4B).

The analog of Weber’s law for facial expressions established in the neuro-perceptual space results remains consistent in the visual space both in neural recordings (Fig. S2A) and in a variant of the psychophysical experiment that regulates expression intensity in the visual space (Fig. S2B). Notably, the same relationship between neural error and facial expression intensity persists for models which incorporate facial motion (Fig. S3).

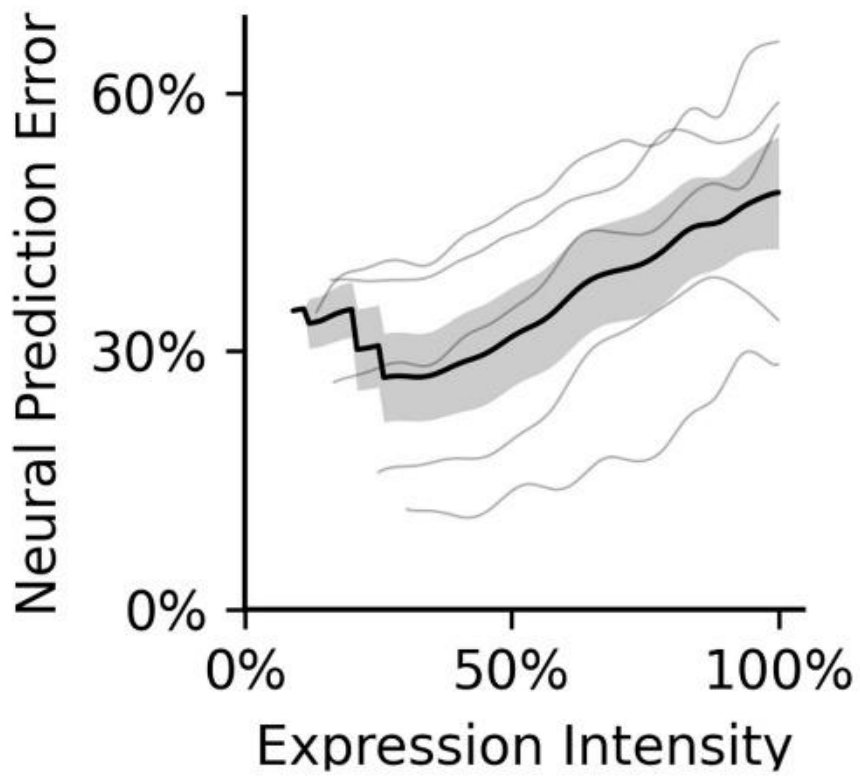

**Fig. S3.** Neural prediction errors for facial expression increased with the intensity of expressions in the neuro-perceptual spaces that incorporated facial motion during fixations.

# Supplementary Discussion

## Prototype Anchored Face Representations and Weber's law

The prototype anchored face coding results were derived from modeling a static representation of fixated faces, i.e., the face representation from the first video frame of each fixation. A Weber's law for facial expressions emerged from the data manifold of the neuro-perceptual spaces of these models and its existence was further validated in a controlled experiment using static images of faces with different expressions. These findings are grounded in established concepts about the neural coding of faces[4–6] such as the idea of a norm face[4] and the notion of facial expressions as deviations from a resting facial appearance[7]. However, facial motion and the considerations that accompany it are largely absent from the theoretical landscape in face processing literature despite a growing interest[8,9] and burgeoning efforts[10–13] to understand how facial motion shapes face perception. For instance, the concept of a person's smile as a deviation from their resting facial expression is intuitive for a snapshot of their face, but it does not represent the difference between the dynamics of someone's resting face breaking into a smile or their smile dissipating into a resting facial expression, in part because the notion of a norm or resting face trajectory remains undefined for dynamic faces. That said, it is notable that our results persist in the neuro-perceptual spaces of models that incorporate facial motion, even though interpreting them to derive meaningful insights remains an open question in this landscape.

## Optimizing for Neuroscientific Discovery Over Reconstruction Quality

The main objective of this study was to unravel the neural code for face perception during natural social interactions in the real world. This goal shaped analytical choices in the computational framework used here, which learned neuro-perceptual spaces optimized such that linear projections of brain activity and facial features were placed close to each other. The analytical choices allowed us to measure non-linearities in how brain activity and facial features corresponded to one another – revealing a Weber's law for the perception of facial expressions that was then probed using a behavioral experiment. The framework also enabled reconstruction of faces and brain activity, although it was not the model's objective.

A prominent approach to improve reconstruction quality and address non-linearities is to use a deep learning model optimized for predictive reconstruction accuracy. However, the geometries of a deep learning network's layers are rarely smooth or systematic enough to be directly interpretable outside of carefully crafted toy examples. This challenge is further amplified when modeling recordings of uncontrolled real world environments where systematic tiling of the stimulus spaces is unlikely and where sample sizes are in the hundreds-to-thousands range instead of millions needed for a deep learning model to generalize well. Altogether, a deep learning approach would make it difficult to make inferences such as Weber's law because the non-linearities of interest would be buried across network layers among a host of transformations that make interpretability challenging. In contrast, a linear model's structure is interpretable and systematic changes in its error profile can reveal the non-linearities it does not capture directly, which opens the door to interpretation and understanding.

Concretely, we measured non-linearities in the data manifold of the shared linear embeddings, i.e., neuro-perceptual spaces, by measuring changes in the error of model fits (neural prediction error) as a function of parametric changes in the stimulus (expression intensity). We observed that neural prediction error increased systematically with increases in expression intensity, in a way that could not result from random noise. We also confirmed that the observed trend of neural prediction error increasing with expression intensity was not due to outliers, i.e., minimizing error for regions of facial expression space where samples were concentrated while incurring higher error in regions with few samples. The trends shown in Fig. 4A, S2A, and S3 are robust to the exclusion of facial expression intensity bins with few samples. Thus, we concluded that the pattern of increasing neural prediction error corresponded to non-linear aspects of the relationship between facial features for expressions and patterns of brain activity, i.e., those not well captured by a linear relationship. In a deep neural network, these non-linearities would be integrated into aspects of the model in ways that can be difficult to extract and leaving it challenging to infer neural tuning properties. Testing specific hypotheses in this manner unearthed the existence of a Weber's law for facial expressions, and the non-linearities that manifest as cone shaped tuning (Fig 4D), serving the purpose of neuroscientific discovery.

## Implications of Model Generalizability for Testing Hypotheses about the Neural Code

The generalizability of neuro-perceptual relationships learned by the model depends upon the diversity and amount of data it is trained with. If the perceptual space is well sampled in training data, then the neuro-perceptual relationships may generalize well, such as the ability to reconstruct facial expressions and motion for individuals absent in training data (Fig. S1B and S1C). Conversely, if regions of the perceptual space were missing from training data, then model performance for predictions on new data from those regions would be around chance, i.e., the model predicts noise, making it difficult to obtain any insights. For instance, a neuro-perceptual space trained on data where a participant interacted with many individuals (>1000) may achieve full out of identity reconstruction of the faces of previously unseen individuals, whereas one trained on interactions with a few individuals (5-10) may not. Thus, testing generalizable hypotheses about the neural code and tuning can only be done robustly in the areas of the model that are well represented by the data acquired, such as for facial expressions and motion in the present study.